\documentclass[12pt]{article}
\usepackage{graphicx}

\begin{document}

 \title{Influence of External Fields and Environment on the Dynamics of
 Phase Qubit-Resonator System }
\author{G.P. Berman $^{1}$\footnote{Corresponding
author: gpb@lanl.gov} and A.A. Chumak$^{1,2}$
\\[3mm]$^1$
 Los Alamos National Laboratory,  MS-B213, Theoretical Division \\ Los Alamos, NM 87545
\\[5mm] $^2$  Institute of Physics of
the National Academy of Sciences\\ pr. Nauki 46, Kiev-28, MSP
03028 Ukraine
\\[3mm]\rule{0cm}{1pt}}
\maketitle \markright{right_head}{LA-UR 11-00280}

\begin{abstract}
We analyze the dynamics of a qubit-resonator system coupled with a
thermal bath and external electromagnetic fields. Using the
evolution equations for the set of Heisenberg operators, that
describe the whole system, we derive an expression for the resonator
field, accounting for the resonator-drive,-bath, and -qubit
interaction. The renormalization of the resonator frequency, caused
by the qubit-resonator interaction, is accounted for. Using
solutions for the resonator field, we derive the equation describing
qubit dynamics. The influence of the qubit evolution during the
measurement time on the fidelity of a single-shot measurement is
studied. The relation between the fidelity and measurement time is
shown explicitly. Also, an expression describing relaxation of the
superposition qubit state towards its stationary value is derived.
The possibility of controlling this state, by varying the amplitude
and frequency of drive, is shown.

\end{abstract}

\section{Introduction}

The possibility of achieving quantum coherence in macroscopic
Josephson junction (JJ) circuits, envisioned by Leggett in the early
1980s \cite{legg1}-\cite{legg3}, was demonstrated experimentally
almost 20 years later by Nakamura {\it et al.} \cite{naka}. Now,
superconductive qubits are promising building blocks for the
realization of a quantum computer. Progress towards quantum
computing depends on the development of measurement schemes and
readout devices. Each readout process requires a finite-time
interval during which the system evolves under the effect of the
measuring device \cite{el1}-\cite{el3} and uncontrollable external
\cite{urb},\cite{intro} or intrinsic \cite{pap}-\cite{shn} noises.
This requires improving the quality of the Josephson contacts and
 effective isolation of qubit systems from the ``electromagnetic
 environment". There is stable progress toward the solution of the first
problems. (See, for example, Ref. \cite{mar}.) The second problem is
fundamental in nature because, for example, it is impossible to
isolate the qubit system from vacuum zero-point oscillations or to
avoid the influence of the measurement device on qubit relaxation or
decoherence.

In this paper, we study the simultaneous influence of external
electromagnetic fields and a thermostat (bath) on the qubit. It is
assumed that this influence is not direct but via a resonator which
is weakly coupled with the qubit. Therefore, the resonator response
on the microwave field depends on the qubit state. In particular, by
measuring the phase of a microwave field reflected from the
resonator, one can perform a nondemolition measurement of the qubit
state. The theoretical description of this scheme and its practical
implementation are elucidated in numerous publications
\cite{el1}-\cite{el3}, \cite{wal4}-\cite{sch7}. At the same time, it
should be emphasized that this kind of measurement results in both
the dephasing of the qubit wave function and partial relaxation
caused by the effects of drive and the thermostat during the
measurement. These effects decrease the measurement fidelity. By
studying the qubit dynamics, we are able to estimate the influence
of drive and bath on the fidelity.

Our further consideration is based on the equations of motion of the
Heisenberg operators. These operators describe the dynamics of both
the qubit and resonator field. The solution of these dynamical
equations will allow us to estimate the upper limit of the
measurement fidelity.

\section{Model}
 We consider the system described by the following Hamiltonian:

\[ H=-{1\over 2}\hbar\omega_q\sigma_z +\hbar\omega_r\bigg
(a^+a+{1\over2}\bigg )+\sum_n \hbar\omega_n\bigg(b^+_nb_n+{1\over
2}\bigg )+\]
\begin{equation}\label{one}
ig\sigma_y\bigg (a^+-a)\bigg)+i\hbar \sum_n f_n\bigg
(b_na^+-b^+_na\bigg )+i\hbar f_0\bigg (ca^+-c^+a\bigg ),
\end{equation}
where the first, second, and third terms in the right side are
Hamiltonians of the noninteracting qubit, the resonator, and the
bath, respectively. The quantity, $\sigma _z$, is the Pauli
operator. We use here the diagonal representation for Hamiltonian of
the isolated qubit. Different types of superconducting qubits, for
example, charge qubit \cite{el1}-\cite{el3} or flux-biased phase
qubit \cite{cla} can be used in practice.

The quantities $a^+,b_n^+,c^+$ are the creation operators of the
corresponding excitations and $a,b_n,c$ are the annihilation
operators. Usually, the drive variables, $c^+$ and $ c$, are
considered to be the classical quantities, $c^*$ and $c$, with given
dependences on time: $c^*,c \sim e^{\pm i\omega_dt}$ in which
$\omega_d$ is the frequency of microwave field.

The fourth, fifth, and sixth terms are the resonator-qubit, -bath,
and -drive ineractions, respectively. The constants $g$, $f_n$,  and
$f_0$, describe the corresponding interaction strengths.  $\omega_q$
is the transition frequency between qubit levels. The quantities,
$\omega_{r}$ and $\omega_{n}$, are the frequencies of the resonator
(cavity) and bath oscillators,  respectively. The resonator
Hamiltonian describes the parallel connection of the local
capacitance, $C$, and the inductance, $L$ ($\omega_r=(LC)^{-1/2}$).
At the same time this lumped-element circuit can represent a
transmission (microstrip) line if we deal with signals whose
characteristic frequencies  are sufficiently close to one of the
eigenfrequencies, $\omega_r$, of the line. More details can be
obtained, for example, in \cite{el1} and \cite{int}.

Similar to papers \cite{wal},\cite{int}, we model the bath as an
infinite set of harmonic oscillators with frequencies, $\omega_n$.
It can be seen from the explicit resonator-bath interaction
Hamiltonian that the presence of the bath is important even at zero
temperature. In this case, only terms describing the annihilation of
resonator excitations and the creation of bath excitations appear.
At finite bath temperatures, there are fluxes of enegy in both
directions: into and out of the bath. There is no unique description
of the effect of the bath on the qubit-resonator system. For
example, the model of a direct qubit-bath interaction is used in
Ref. \cite{joha}.

The resonator-bath as well as the resonator-drive interactions
affect the qubit state due to the qubit-resonator interaction. This
situation is very similar to the case considered in \cite{urb} in
which semiclassical noise of the biased current causes qubit
decoherence and relaxation.

\section{Evolution of the resonator field}

Using the Heisenberg representation for the operators in Eq.
\ref{one}, we can express the time derivative of $a$ as
\begin{equation}\label{two}
\dot{a}=\frac 1{i\hbar}[a,H]=-i\omega_ra+\frac
g\hbar\sigma_y+\sum_nf_nb_n+f_0c.
\end{equation}
Here, the dependences of $\sigma_y$ and $b_n$ on time are not yet
specified. To determine the explicit dependences of $\sigma_y(t)$
and $b_n(t)$, we will use an iterative procedure that assumes that
all interaction parameters are small quantities. As in Eq. \ref{two}
we have
\begin{equation}\label{thr}
\dot{\sigma_y}=-\omega_q\sigma_x,
\end{equation}
\[\dot{\sigma_x}=\omega_q\sigma_y+i2\frac g\hbar\sigma_z(a^+-a).\]
From these equations we can obtain relationships for the more
convenient variables $\sigma_\pm\equiv\frac 12(\sigma_x\pm
i\sigma_y)$:
\begin{equation}\label{fou}
\dot{\sigma}_\pm=\mp i\omega_q\sigma_\pm+i\frac g\hbar
\sigma_z(a^+-a).
\end{equation}
 Eqs. \ref{fou} can be rewritten in the
equivalent form as
\begin{equation}\label{fiv}
\sigma_\pm (t)=\sigma_\pm (t_0)e^{\mp i\omega_q(t-t_0)}+i\frac
g\hbar \int_{t_0}^tdt^\prime e^{\mp
i\omega_q(t-t^\prime)}[\sigma_z(a^+-a)]_{t^\prime}.
\end{equation}
Further analysis is facilitated by assuming the qubit-resonator
detuning is small:
\[|\omega_{qr}|<<\omega_q,\omega_r,\quad \omega_{qr}\equiv\omega_q-\omega_r.\]
(We assume this inequality is satisfied throughout the paper.) At
the same time the detuning should be large compared to the
interaction parameter $g$ :
\begin{equation}\label{six}
 \frac g{\hbar |\omega_{qr}|}<<1.
\end{equation}
In many papers (see, for example, Refs. \cite{el1}-\cite {el3}), the
quantity in the left side of Eq. \ref{six} is also considered  as
small parameter. In this case, the dynamics of the system can be
analysed using a simpler (renormalized) Hamiltonian.

Considering $\sigma_z(t^\prime)a^+(t^\prime)e^{-i\omega_rt^\prime}$
and $\sigma_z(t^\prime)a(t^\prime)e^{i\omega_rt^ \prime}$ as slowly
varying functions of $t^\prime$, we can integrate over $t^\prime$ in
Eq. \ref{fiv}. The result is

\begin{equation}\label{sev}
\sigma_+ (t)=\tilde{\sigma}_+ (t)-\frac g\hbar \frac
{\sigma_z(t)a(t)}{\omega_{qr}},
\end{equation}
\[\sigma_- (t)=\tilde{\sigma}_- (t)-\frac g\hbar \frac
{\sigma_z(t)a^+(t)}{\omega_{qr}},\] where $\tilde{\sigma}_\pm
=\sigma_\pm(t_0)e^{\mp i\omega_q(t-t_0)}$. In the course of
integration, we have assumed that $t-t_0\rightarrow \infty$.

Using Eqs. \ref{sev} we can express $\sigma_y(t)$, which appears in
Eq. \ref{two}, in terms of $\tilde{\sigma}_\pm (t),\sigma_z(t),
a^+(t),$ and $a(t)$. The dependences of $b^+_n,b_n$ on time can be
also expressed in terms of $a^+$ and $a$. As in Eqs. \ref{fou}, we
have
\begin{equation}\label{eig}
\dot{b_n}=-i\omega_nb_n-f_na,
\end{equation}
or
\begin{equation}\label{nin}
b_n(t)=\tilde{b}_n(t)-f_n\int_{t_0}^tdt^\prime
e^{-i\omega_n(t-t^\prime)}a(t^\prime ),
\end{equation}
where $\tilde{b}_n(t)=b_n(t_0)e^{-i\omega_n(t-t_0)}$.

 Multiplying both sides of Eq. \ref{nin} by $f_n$ and summing
over $n$ we obtain
\begin{equation}\label{ten}
\sum_nf_nb_n(t)=\sum_nf_n\tilde{b}_n(t)-\int_{t_0}^tdt^\prime
\sum_nf_n^2e^{-i\omega_n(t-t^\prime)}a(t^\prime ).
\end{equation}
Using a simple approximation for the sum in the integrand of Eq.
\ref{ten}
\begin{equation}\label{ele}
\sum_nf_n^2e^{-i\omega_n(t-t^\prime)}=\kappa \delta(t-t^\prime),
\end{equation}
(see Refs. \cite{wal},\cite{int}), we can rewrite Eq. \ref{ten} in a
simple form:
\begin{equation}\label{tve}
\sum_nf_nb_n(t)=\sum_nf_n\tilde{b}_n(t)-\frac \kappa 2 a(t).
\end{equation}
Then using Eqs. \ref{sev} and \ref{tve}, the equation for $a(t)$
reduces to
\begin{equation}\label{thi}
\bigg [\partial _t+i\bigg (\omega_r- \chi\sigma_z\bigg )+
\frac{\kappa}{2}\bigg ]a=f_0c+\sum_nf_n\tilde{b}_n- i\frac
{g}{\hbar}\bigg(\tilde{\sigma}_+-\tilde{\sigma}_-\bigg
)-i\chi\sigma_za^+,
\end{equation}
where  $\chi \equiv {g^2}/(\hbar ^2\omega_{qr})$.

The quantity, $\omega_r-\chi\sigma_z$, is the resonator frequency
renormalized by the qubit-resonator interaction. This
renormalization effect can be derived from the Jaynes-Cummings
Hamiltonian using a unitary transformation, assuming that $g/(\hbar
\omega_{qr})$ is a small parameter.

The parameter, $\kappa/2$, appearing in the bath-resonator
interaction, is an important characteristic of the resonator. It can
be seen from the structure of Eq. \ref{thi} that $\kappa/2$
describes the field dissipation caused by this interaction. The
ratio $\omega_r/\kappa$ is the resonator quality factor, $Q$.
Usually, high-$Q$ resonators are used for qubit measurements.

The solution of Eq. \ref{thi} (in which the influence of the initial
condition or transient stage is ignored) is given by:
\begin{equation}\label{fort}
 a(t)=\frac {if_0c(t)}{\tilde{\omega}_{dr}+i\kappa/2}+
 \sum_n\frac
 {if_n\tilde{b}_n(t)}{\tilde{\omega}_{nr}+i\kappa/2}+\frac
 g\hbar\frac {\tilde{\sigma}_+(t)}{\tilde{\omega}_{qr}+i\kappa/2},
\end{equation}
where $\tilde{\omega}_{ir}\equiv \omega_i-\omega_r+\chi\sigma_z,
i=d,n,q$. In the course of solution of Eq. \ref{thi}, explicit
dependences of $c,\tilde{b}_n,\tilde{\sigma}_\pm$ on $t$ were used.
Also, the contribution of terms with $\sigma_-$ and $a^+$ was
neglected. This approximation is accurate when $\chi , (g/\hbar
)<<\omega_r$.

The value for $a^+$ can be obtained from Eq. \ref{fort} using
hermitian conjugation. It is given by
\begin{equation}\label{fift}
 a^+(t)=\frac {if_0^*c^+(t)}{\tilde{\omega}_{rd}+i\kappa/2}+
 \sum_n\frac
 {if_n^*\tilde{b}^+_n(t)}{\tilde{\omega}_{rn}+i\kappa/2}-
 \frac g\hbar \frac
 {\tilde{\sigma}_-(t)}{\tilde{\omega}_{rq}+i\kappa/2}.
 \end{equation}

Using Eqs. \ref{fort}, \ref{fift}, and \ref{ele}, we can easily show
that the standard commutation relations between operators $a$ and
$a^+$ are fulfilled with an accuracy valid up to a small value of
the order $g^2/(\hbar ^2\omega_{qr}^2)$ if the drive variables are
considered as classical quantities. The deviation of $[a,a^+]$ from
unity is within the accuracy of perturbation procedure used here.

It follows from Eqs. \ref{fort} and \ref{fift} that the effect of
bath is represented by the second (``noise") terms and the imaginary
summand, $i\kappa/2$, in the denominators. The effect of drive on
the resonator field critically depends on the detuning
$\tilde{\omega}_{dr}=\omega_d-\omega_r+\chi \sigma_z$. The field
amplitude and the photon number in the cavity, $n_r=a^+a$, are the
largest when the detuning is of the order of $k/2$. If the drive
frequency is fixed, the detuning depends on the qubit state. For the
upper and lower states, the resonant conditions can be very
different when $\chi>\kappa/2$. This circumstance is commonly used
for measuring qubit states by means of microwave fields (see, for
example, Ref. \cite{sch5}).

To estimate the importance of the different terms in Eqs. \ref{fort}
and \ref{fift},
 we will calculate the average (over bath variables) photon number in the
 resonator, $n_r^b$, considering the qubit to be in the excited
 ($\sigma_z=- 1$) or ground ($\sigma_z= 1$) state. (The frequency,
 $\tilde{\omega}_{dr}$, is equal to
 $\omega_d-
 \omega_r\mp\chi$ for $\sigma_z=\mp 1$,
 respectively.)
 Using Eq. \ref{ele}
 and the relationship
\begin{equation}\label{sixt}
 \sigma_-\sigma_+=\frac {1}{2}(1-\sigma_z),
\end{equation}
we obtain
\begin{equation}\label{seve}
 n_r^b=\frac {|f_0c|^2}{\tilde{\omega}_{dr}^2+\kappa^2/4}+\langle
 b^+_nb_n\rangle _
 {\omega_n=\omega_r}+\frac {g^2}{2\hbar^2}
 \frac {1-\sigma_z}{\tilde{\omega}_{qr}^2},
\end{equation}
where correlations between different bath modes were ignored
($\langle b^+_nb_{n^\prime}\rangle\sim\delta_{n,n^\prime}$). In the
course of derivation of the second term in Eq. \ref{seve}, we have
considered that the average $\langle b^+_nb_n\rangle$ depends on $n$
via $\omega_n$ only. Then using Eq. \ref{ele} we were able to sum up
over bath modes as:
\[\sum_n\frac {f_n^2\langle b^+_nb_n\rangle}{\tilde{\omega}_{nr}^2+
\kappa^2/4}=\sum_n\int_{-\infty}^\infty d\omega \delta (\omega-
\tilde{\omega}_{nr})\frac {f_n^2\langle
b^+_nb_n\rangle|_{\omega_n=\omega+\omega_r-\chi\sigma_z}}{\omega^2+
\kappa^2/4}\approx\] \[\langle
b^+_nb_n\rangle|_{\omega_n=\omega_r}\int_{-\infty}^ \infty \frac
{d\omega}{\omega^2+\kappa^2/4} \int_{-\infty}^ \infty \frac
{d\tau}{2\pi}
\sum_nf_n^2e^{i(\omega-\tilde{\omega}_{nr})\tau}=\langle
b^+_nb_n\rangle|_{\omega_n=\omega_r}.\]

 The remarkable peculiarity of Eq. \ref{seve} is
that the direct contribution of bath does not depend on the
interaction constant, $f_n$. This is in contrast to Eqs. \ref{fort}
and \ref{fift} for the resonator field.

For thermal equilibrium, the average occupancy of the bath modes is
given by the Bose-Einstain distribution function $n_{BE}$
\begin{equation}\label{eigh}
 \langle b^+_nb_n\rangle =n_{BE}(\omega _n)\equiv \bigg (e^{\frac {\hbar \omega_n}
 {k_BT}}
 -1\bigg )^{-1}
\end{equation}
Therefore, in the absence of the drive and the qubit, the number of
photons in the resonator is equal to that in the corresponding bath
mode. In other words, the bath and resonator temperatures are equal
in this (equilibrium) case.

Eq. \ref{seve} shows explicitly when the driving field dominates the
noisy influence of the bath. Besides that, it follows from Eq.
\ref{seve} that the qubit ``delivers" an almost negligible portion
of photons to the resonator (the last term in Eq. \ref{seve}) even
in the most favorable case, $\sigma_z=-1$. The physical reason for
this is in the qubit-resonator detuning. The detuning decreases the
probability of qubit excitations to ``penetrate" into the cavity.

Eqs. \ref{fort} and \ref{fift} can be used to study the fluctuations
of photon numbers in the resonator. These fluctuations are
responsible for qubit decoherence (more details can be obtained, for
example, in Ref. \cite{el2}).

In the next Section, we will use Eqs. \ref{fort} and \ref{fift} to
describe the qubit dynamics. Drive- and thermostat-induced
variations of the qubit state during the measurement time will be
studied. These variations are responsible for reducing the
measurement fidelity of the qubit state.

\section{Qubit evolution}

The time variation of qubit states occupancies can be expressed in
terms of the average value of the operator $\sigma_z$. When the
qubit is in the state $\psi (t)=\alpha (t)|0\rangle+\beta
(t)|1\rangle$, the average value of $\sigma_z$ is:
\[ \langle \sigma_z\rangle\rangle _t=\langle \psi (t)|\sigma_z|\psi (t)\rangle =|\alpha
(t)|^2-|\beta(t)|^2. \] If we know $\langle \sigma_z\rangle _t$, the
occupancies of the levels can be obtained from:
 \[|\alpha (t)|^2=\frac
12(1+\langle \sigma_z\rangle _t),\quad |\beta(t)|^2=\frac
12(1-\langle \sigma_z\rangle _t).\] We will replace $\langle
\sigma_z\rangle _t$ by $\langle \sigma_z(t)\rangle$, where
$\sigma_z(t)$ is defined in the Heisenberg representation and
averaging is over the initial state, $\psi (t_0)$. Therefore, the
evolution of the occupancies can be obtained from the temporal
dependence of the operator $\sigma_z(t)$.

To obtain $\sigma_z(t)$, we use, as previously, the equations of
motion for the corresponding operators. The time variation for
$\sigma_z(t)$ is:
\begin{equation}\label{nine}
 \dot{\sigma}_z=\frac 1{i\hbar}[\sigma_z,ig\sigma_y(a^+-a)].
\end{equation}
We can express $\sigma_y(a^+-a)$ as $-i(\sigma_+a^++\sigma_-a)$  in
the spirit of rotating-wave approximation. Then Eq. \ref{nine}
reduces to
\begin{equation}\label{twen}
 \dot{\sigma}_z\approx-i\frac {2g}\hbar(\sigma_+a^+-a\sigma_-).
\end{equation}
In the next step, we will express the right-hand side of Eq.
\ref{twen} in terms of $\sigma_z(t)$. Similar to the consideration
in the previous Section, we use the equation of motion for the
operators $\sigma_+a^+,\,\sigma_-a$. Thus we have
\begin{equation}\label{twon}
 (\partial_t+i\omega_{qr})\sigma_+a^+=i\frac {g}\hbar\sigma_za^+(a^+-a)+
 i\frac {g}{2\hbar}(1-\sigma_z)+\sigma_+\bigg (f_0c^++\sum_nf_nb^+_n\bigg).
\end{equation}
Using Eq. \ref{tve} and neglecting the term containing $a^+a^+$, we
can rewrite Eq. \ref{twon} as
\begin{equation}\label{twtw}
 \bigg(\partial_t+i\omega_{qr}+\frac \kappa 2\bigg)\sigma_+a^+=-i\frac {g}\hbar
 \sigma_z\bigg(a^+a+\frac 12\bigg)+
 i\frac {g}{2\hbar}+\sigma_+\bigg (f_0c^++\sum_nf_n\tilde{b}^+_n\bigg ).
\end{equation}
Using  Eqs. \ref{sev} and \ref{fort}, we obtain from Eq. \ref{twtw}

\[ \bigg(\partial_t+i\omega_{qr}+\frac \kappa
2\bigg)\sigma_+a^+=\tilde{\sigma}_+
 \bigg(f_0c^++\sum_nf_n\tilde{b}_n^+\bigg)-
i\frac g\hbar  \bigg[\sigma_z\bigg(a^+a+\frac 12\bigg)-\frac
12\bigg]+\]
\begin{equation}\label{twth}
\frac {g\sigma _z}{\hbar \omega _{qr}}\bigg (\frac
{f_0c}{\tilde{\omega}_{dr}+i\kappa /2}+\sum_n\frac
{f_n\tilde{b}_n}{\tilde{\omega}_{nr}+i\kappa /2}\bigg)\bigg
(f_0c^++\sum_{n^\prime} f_{n^\prime} \tilde{b}_{n^\prime}^+\bigg).
\end{equation}
Let us average both sides of Eq. \ref{twth} over the thermostat
variables. Then again considering $\sigma_z$ as a slowly varying
function of time, we can easily obtain
$\langle\sigma_+a^+\rangle_{bath}$ from Eq. \ref{twth} in the form

\[\langle \sigma_+a^+\rangle _{bath}=\frac {ig\bigg(e^{(-i\omega_{qr}-\kappa
/2)t}-1\bigg)}{\hbar (i\omega_{qr}+\kappa/2)} \bigg
\{\bigg[\sigma_z\bigg(n_r^b+\frac 12\bigg)-\frac 12\bigg]+\]
\begin{equation}\label{twfo}
 i\frac {\sigma_z}{\omega_{qr}}
\bigg (\frac {|f_0c|^2}{\tilde{\omega}_{dr}+i\kappa/2}+\sum_{n}
\frac {f^2_n \langle
b_nb_n^+\rangle}{\tilde{\omega}_{nr}+i\kappa/2}\bigg)\bigg\},
\end{equation}
where we have neglected the initial (at $t=0$) correlations of the
operator $\sigma_+$ with the operators $a^+, c^+$. The value of
$n_r^b$ is given by Eq. \ref{seve} in which the last term,
representing the contribution of qubit, can be omitted due to its
small value.

The expression for $\langle a\sigma_z\rangle _{bath}$ can be derived
in a similar manner. Then the equation for $\sigma_z$ is:
\[\dot{\sigma}_z=\frac {2g^2}{\hbar^2\omega_{qr}^2}\bigg\{\bigg [2
\omega_{qr}sin(\omega_{qr}t)e^{-\frac {\kappa} 2t}+\kappa
\bigg(1-cos(\omega_{qr}t)e^{-\frac \kappa 2t}\bigg)\bigg ]
\bigg[\frac 12-\sigma_z\bigg(n_r^b+\frac 12\bigg) \bigg]-
\]
\begin{equation}\label{twfi}
\sigma_z\frac {|f_0c|^2}{\tilde{\omega}^2_{dr}+\kappa^2/4}\bigg[2
\tilde{\omega} _{dr}\bigg(1-cos(\omega_{qr}t)e^{-\frac \kappa
2t}\bigg)+\kappa sin(\omega_{qr}t)e^{-\frac \kappa 2t}\bigg] \bigg
\}.
\end{equation}
Because $\omega_{qr}>>\kappa/2$, we can omit in Eq. \ref{twfi} the
oscillating terms that are proportional to $\kappa$ and
$\tilde{\omega}_{dr}$. Thus, the rate equation reduces to
\begin{equation}\label{twsi}
\dot{\sigma}_z=\frac {2g^2}{\hbar^2\omega_{qr}^2}\bigg\{\bigg [2
\omega_{qr}sin(\omega_{qr}t)e^{-\frac {\kappa} 2t}+\kappa \bigg ]
\bigg[\frac 12-\sigma_z\bigg(n_r^b+\frac 12\bigg)
\bigg]-\sigma_z\frac {2 \tilde{\omega}
_{dr}|f_0c|^2}{\tilde{\omega}^2_{dr}+\kappa^2/4}\bigg \}.
\end{equation}

 In the absence of
drive ($f_0=0$), Eq. \ref{twsi} describes small-amplitude
thermostat-induced Rabi oscillations with  frequency $\omega_{qr}$.
These oscillations decay during the coherence time, $\sim 2/\kappa$,
of the resonator field. The oscillations are accompanied by a slow
qubit relaxation to the stationary value
\begin{equation}\label{addi}
\sigma_{st}=\bigg(1+2\langle b^+_nb_n\rangle
|_{\omega_n=\omega_r}\bigg)^{-1}.
\end{equation}
(See, for example, Ref. \cite{gar}.) The amplitude of oscillations
depends on: the occupancies of the bath states, $\langle
b^+_nb_n\rangle |_{\omega_n=\omega_r}$, the cavity losses, $\kappa
/2$, and the qubit-resonator detuning, $\omega_{qr}$. In the case of
an equilibrium bath, $\sigma_{st}$ coincides with the equilibrium
value known in the literature. It can be derived in an alternative
manner using the density matrix  formalism.

The case of $f_0\neq0$ is more interesting in view of the
possibility of using the drive to measure (control) the qubit. The
preferred setup is realized if the microwave field is in  resonance
with the cavity-qubit system in which the qubit is in a given
eigenstate (for example, in the excited state:
$\omega_d=\omega_r+\chi$). (This is the case for which the last term
in braces of Eq. \ref{twsi} vanishes.) The measuring device can
resolve the resonator frequencies $\omega_r\pm\chi$, corresponding
to the different eigenstates of the qubit, only if the measurement
time, $\tau$, is greater than $(2\chi )^{-1}$. At the same time,
$\tau$ should be as small as possible to provide rapid control.
Moreover, during the measurement time, the qubit evolves as
described by the rate equation (\ref{twsi}), thus decreasing the
measurement fidelity $F$.
 As a consequence of the measurement, an arbitrary superposition
 qubit state
 collapses to one of the
eigenstates with $\sigma_z=\pm1$. The characteristic time of the
collapse (defined here as the decoherence time) is assumed to be
considerably shorter than $\tau$. Therefore, the fidelity of the
measurement, $F$, for the qubit in the post-collapse state, $|1>$,
is defined as
\begin{equation}\label{twse}
F={1\over\tau}\int_0^\tau dt|\beta(t)|={1\over\tau}\int_0^\tau
dt\bigg(\frac{1-\sigma_z}2\bigg)^{1/2}\approx
1-\frac{g^2(n_r^b+1)}{\hbar^2\omega_{qr}^2}\bigg(1-\frac{sin(\omega_{qr}\tau)}
{\omega_{qr}\tau}\bigg).
\end{equation}
For simplicity, we have considered the case  $\tau <2/\kappa$ in
which $\sigma_z$ is given by
\[\sigma_z(t)\approx -1+\frac{8g^2(n_r^b+1)}{\hbar^2\omega_{qr}^2}sin^2
\bigg(\frac{\omega_{qr}t}2\bigg).\] Taking into account
$\omega_{qr}\tau>>1$, we have
\begin{equation}\label{twei}
F\approx 1-\frac{g^2(n_r^b+1)}{\hbar^2\omega_{qr}^2}.
\end{equation}
This formula illustrates the effect of the Rabi oscillations,
generated
 by the external drive and bath, on the fidelity. In contrast to $\sigma_z(t)$,
 the fidelity does not display the oscillating behavior. This is because
 the value of   $\sigma_z$ is averaged
 over the interval $\tau >>|\omega_{qr}|^{-1}$. For longer
measurement times,  $\tau >2/\kappa$ , all terms in Eq. \ref{twsi}
should be used to calculate $\sigma_z(t)$ and $F(\tau)$.

It seems from Eq. \ref{twei} that the fidelity can be improved for
smaller interaction parameters. But decreasing $g$ will decrease
$\chi =g^2/(\hbar ^2\omega_{qr})$. In view of the inequality $\tau
>(2\chi)^{-1}$ and using Eq. \ref{twei}, we obtain
\begin{equation}\label{a}
\tau >\frac{n_r^b+1}{2(1-F)\omega_{qr}}.
\end{equation}
It follows from Eq. \ref{a} that as  $F$ approaches unity, the
measurement time should be increased.

 We have described theoretically a single-shot measurement
 in which the initial qubit
 was assumed to be in  either the excited or the ground state. The evolution of
 the superposed state
can be investigated experimentally by the repetition of many
single-shot measurements. The rate equation for $\langle
\sigma_z(t)\rangle$ [the averaging is over the initial state,
$\psi(t=0)$], corresponding to this kind of the experiment, can be
derived in a manner similar to Eq. \ref{twsi}. It is given by
\begin{equation}\label{twni}
\langle \dot{\sigma}_z\rangle =\frac{\kappa
g^2}{\hbar^2\omega_{qr}^2}\bigg[-\langle{\sigma}_z\rangle
\big(\varphi^++\varphi^-\big)+\varphi^--\varphi^++1\bigg] ,
\end{equation}
 where the initial (oscillatory) stage of evolution is ignored;
\begin{equation}\label{ad}
\varphi^\pm \equiv \varphi (\sigma_z =\pm1),\quad \varphi
(\sigma_z)=n_r^b+\frac12+\frac {2\tilde{\omega}_{dr}}\kappa\frac
{|f_0c|^2} {\tilde{\omega}_{dr}^2+\kappa^2/4}  .
\end{equation}
In the derivation of Eq. \ref{twni} the following identities were
used:
\[\varphi(\sigma_z)=\frac
12[(1+\sigma_z)\varphi^++(1-\sigma_z)\varphi^-].\] and
\begin{equation}\label{thirt}
\sigma_z \varphi(\sigma_z)=\frac
12[(1+\sigma_z)\varphi^+-(1-\sigma_z)\varphi^-].
\end{equation}
The solution of Eq. \ref{twni} is given by
\begin{equation}\label{thiro}
\langle \sigma_z(t)\rangle =\sigma_{st}+\bigg(\langle
\sigma_z(t=0)\rangle -\sigma_{st}\bigg)e^{-\gamma t}  ,
\end{equation}
in which the stationary value, $\sigma_{st}$, and the relaxation
constant, $\gamma$, are given by
\begin{equation}\label{thitw}
\sigma_{st}=\frac {\varphi^--\varphi^++1}{\varphi^-+\varphi^+}
,\quad \gamma
=\frac{kg^2\big(\varphi^-+\varphi^+\big)}{\hbar^2\omega_{qr}^2}.
\end{equation}
In the limiting case $f_0=0$, $\sigma_{st}$ reduces to the previous
result given by Eq. \ref{addi}. In the  limit of dominating drive in
Eq. \ref{seve} ($n_r^b>>1$ for the resonance conditions,
$\omega_d=\omega_r\pm\chi$), the qubit relaxes to  the ground or
excited states, respectively. In the case of large resonator-drive
detuning, $|\omega_{dr}|>>\chi$, the last term in Eq. \ref{ad} can
 dominate. Then $\langle\sigma_{z}\rangle$ relaxes to zero.

\section{Conclusion}

We have described a qubit-resonator system with an external drive
and thermostat. Our consideration is based on the equations of
motion of the operators in the Heisenberg representation. This is in
contrast to the widely used density matrix  approach. Considering
the resonator-bath, resonator-drive, and resonator-qubit
interactions as weak perturbations, we have derived  expressions for
the resonator field including the renormalization of the resonator
frequency caused by the qubit-resonator interaction. A weak
qubit-environment interaction is a necessary condition for reliable
isolation of the qubit from the ``external wold.

Also, we have derived the rate equation for the qubit variable,
$\sigma_z$, describing the occupancies of the qubit levels. The
solution of this rate equation enables us to calculate the
measurement fidelity, $F$, and to determine the dependence of $F$ on
the measurement time, $\tau$: increasing fidelity requires
increasing the measurement time. (See Eq. \ref{a}.) Both quantities
are very important parameters in view of the practical
implementations of qubits. Therefore, the optimal choice of $F$ and
$\tau$ should be carried out considering their interdependency given
by inequality \ref{a}.

The qubit relaxation, caused by the interactions with the bath and
the drive, can be used to control the final qubit state. In
particular, it follows from Eqs. \ref{twni}-\ref{thitw} that, by
varying the frequency and amplitude of the drive as well as the
interaction time, $t$, we can get a qubit with a predetermined
probabilities to be in the ground or excited state.

\section{Acknowledgment}
We are grateful to V.I. Tsifrinovich, D.I. Kamenev, and D. Kinion
for useful discussions. This work was carried out under the auspices
of the National Nuclear Security Administration of the U.S.
Department of Energy at Los Alamos National Laboratory under
Contract No. DE-AC52-06NA25396 and by Lawrence Livermore National
Laboratory under Contract DE-AC52- 07NA27344, and was funded by the
Office of the Director of National Intelligence (ODNI), and
Intelligence Advanced Research Projects Activity (IARPA). All
statements of fact, opinion or conclusions contained herein are
those of the authors and should not be construed as representing the
official views or policies of IARPA, the ODNI, or the U.S.
Government.

\newpage \parindent 0 cm \parskip=5mm


\end{document}